\documentclass{article}

\PassOptionsToPackage{numbers,sort&compress}{natbib}
\usepackage[preprint]{style} 

\usepackage[utf8]{inputenc} 
\usepackage[T1]{fontenc}    
\usepackage{hyperref}       
\usepackage{url}            
\usepackage{booktabs}       
\usepackage{amsfonts}       
\usepackage{nicefrac}       
\usepackage{microtype}      
\usepackage{amsmath,amsthm,amsfonts,amssymb,amscd}
\usepackage{mathtools}
\usepackage{lastpage}
\usepackage{enumerate}
\usepackage{fancyhdr}
\usepackage{mathrsfs}
\usepackage[x11names]{xcolor}
\usepackage{graphicx}
\usepackage{listings}
\usepackage{subcaption}
\usepackage[english]{babel}
\usepackage{apxproof}
\usepackage{thmtools}
\usepackage{thm-restate}
\usepackage{enumitem}
\usepackage{float}
\usepackage{bm}
\usepackage{wrapfig}
\usepackage{algorithm}
\usepackage{algpseudocode}

\usepackage{sidecap}
\usepackage{siunitx}

\theoremstyle{definition}

\definecolor{myGreen}{rgb}{0,0.5,0}

\title{Curved Micro-Electrode Arrays}

\author{
  Markus Meister\\
  Division of Biology and Biological Engineering\\
  California Institute of Technology\\
  \texttt{meister@caltech.edu}
}

\begin{document}

\maketitle

\begin{abstract}
Multi-electrode arrays serve to record electrical signals of many neurons in the brain simultaneously \cite{luanRecentAdvancesElectrical2020a,steinmetzNeuropixelsMiniaturizedHighdensity2021}. For most of the past century, electrodes that penetrate brain tissue have had exactly one shape: a straight needle \cite{renshawActivityIsocortexHippocampus1940}. Certainly this was a good starting choice at the time, but there is no reason to think that a straight line would be the optimal shape in all Neuroscience applications. Here I argue that, in fact, a wide variety of curved shapes is equally practical: all possible helices. I discuss the manufacture and manipulation of such devices, and illustrate a few use cases where they will likely outperform conventional needles. With some collective action from the research community, curved arrays could be manufactured and distributed at low cost.
\end{abstract}

\section{Circular arrays}
\begin{figure}
  \centering
    \includegraphics[width=0.75\linewidth]{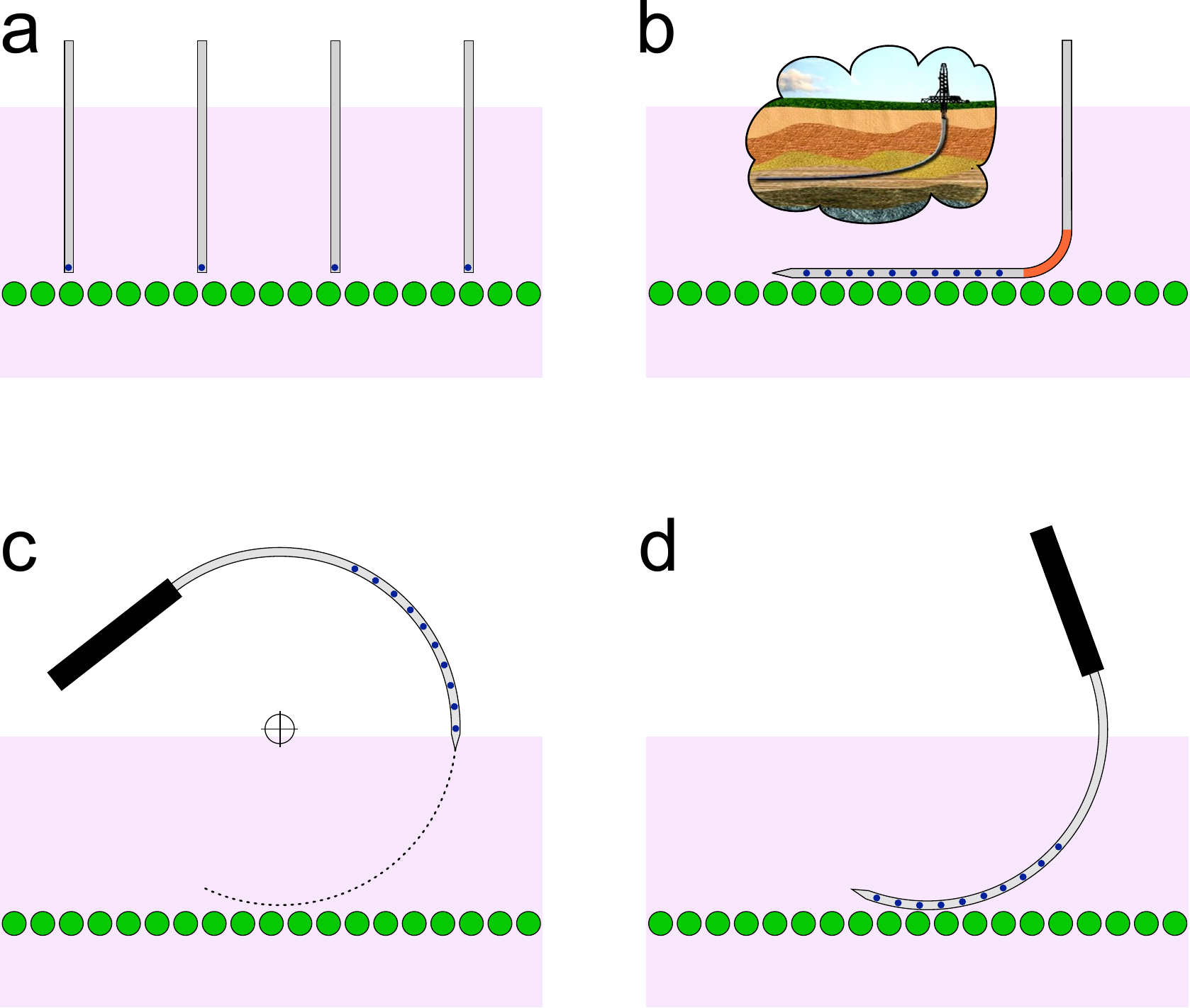}
    \caption{\textbf{How to record from neurons arranged in a plane.} \textbf{a:} Schematic of hippocampal pyramidal cells (green circles) arranged in a layer deep in the brain (pink). Conventional approach: insert several tetrodes, each with recording sites (blue) only at the tip. \textbf{b:} Ideally the electrode would bend when it reaches the desired depth, as in oil drilling with steerable boring heads. As the electrode advances, the bent region (red) would have to propagate up the shaft without exerting lateral forces. \textbf{c:} A stiff multielectrode array shaped as a circular arc. The array can be inserted by rotating the shank and its connector (black) around the axis of the circle (cross). \textbf{d:} Final arrangement in which the array follows the plane of cell bodies for a considerable distance. Drawing not to scale.}
\label{fig:circular}
\end{figure}

What might be an application that calls for curved electrodes? A popular target for neural recording are the pyramidal cells in area CA1 of the hippocampus \cite{okeefeReviewHippocampalPlace1979,solteszCA1PyramidalCell2018}.
\footnote{Incidentally this includes the first historical use of penetrating microelectrodes \cite{renshawActivityIsocortexHippocampus1940}.}
The firing of these neurons represents where the animal is located in its environment. The code is sparse: In any given environment only a small fraction of these neurons are active. In order to understand how the brain maps space, one therefore needs to record densely from many neurons in this population. Anatomically the cells are arranged neatly in a plane, roughly parallel to the brain's surface (Fig \ref{fig:circular}a). If one sinks a linear electrode array into the brain, it will intersect this plane at exactly one point, and thus record only one or a few nearby neurons. The currently accepted method is to sink many straight wire bundles (called tetrodes) vertically into the brain so they come to rest just above the pyramidal cells (Fig \ref{fig:circular}a) \cite{lubenovHippocampalThetaOscillations2009,redishIndependenceFiringCorrelates2001}. Of course each of these tetrodes carves a cylinder of destruction through the overlying neural circuitry of cortex and hippocampus. Furthermore, mechanical constraints force the tetrodes several 100 µm apart, so at best they sample the pyramidal cell layer only sparsely. 

It would be preferable if a single electrode array could be induced to turn at a right angle just above the pyramidal cell layer and then proceed horizontally, such that electrodes all along its shaft can sample many neurons (Fig \ref{fig:circular}b). Similar thoughts must have occurred in the petrochemical industry, leading to the invention of the steerable drill head, which revolutionized oil and gas exploration by allowing horizontal drilling. Unfortunately the scale and material properties of neural tissue are rather different, and a solution analogous to the steerable drill head seems impractical. In particular, as the electrode is advanced by pushing from above, the bend region (Fig \ref{fig:circular}b) would need to coordinately travel up along the shaft so there is never any sideways force on the bottom portion that would make it slice through the tissue.

Fortunately a much simpler solution is within reach: a stiff array that is pre-bent in a circular shape (Fig \ref{fig:circular}c). After rotating such an electrode array into the brain, a new recording geometry results. If the diameter of the circle and the insertion point are chosen appropriately then a good portion of the array will run parallel to the plane of the target neurons (Fig \ref{fig:circular}d). In this way a single device can record many more neurons than under the conventional approach, and thus the damage to overlying brain tissue is reduced. Furthermore, at least along the path of this array, one can record densely from every pyramidal cell within reach.

\begin{figure}
  \centering
    \includegraphics[width=0.75\linewidth]{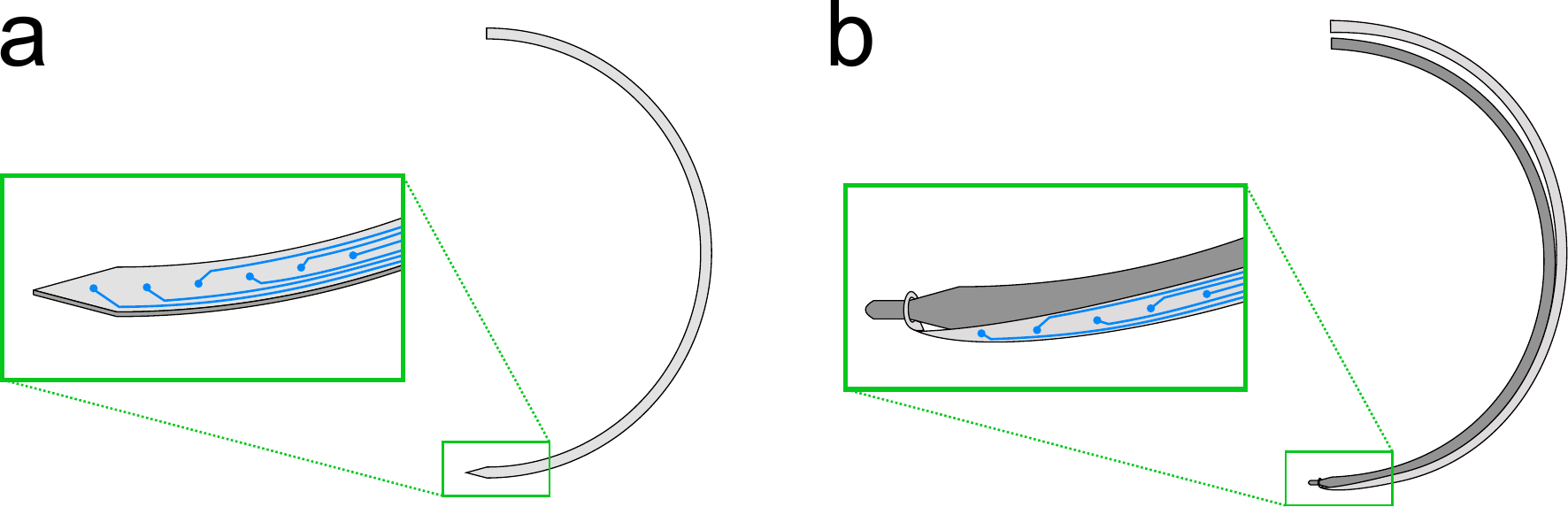}
    \caption{\textbf{How to manufacture a circular array.} \textbf{a:} A silicon array can be manufactured by conventional e-beam lithography using existing designs for straight arrays and curving all the lines. \textbf{b:} A flexible polymer array can be inserted with a circular needle acting as shuttle. The polymer thread is attached to the needle by a loop or with dissolvable glue. When the needle withdraws, only the polymer thread is left behind.}
\label{fig:construction}
\end{figure}

One can manufacture such a circular electrode array in silicon by the same methods used for linear electrode arrays \cite{duMultiplexedHighDensity2011,moralopezNeuralProbe9662017}. The array will take the shape of a thin annulus. All the electrodes and wires are patterned onto the flat side of the annulus. The lithography beam simply needs to draw curves instead of straight lines (Fig \ref{fig:construction}a). 

Insertion of such a curved array into the brain proceeds by rotation, rather than via the usual linear translation stage (Fig \ref{fig:circular}c-d). If the radius of curvature is large compared to the thickness of the array, then the motion and forces on the tissue are purely tangential to the shaft. Thus the tissue will respond as if a straight array were advanced on a straight trajectory. However the forces on the array are different from the linear case: At the time of insertion the curved array will experience a large bending moment at the base of the shaft (Fig \ref{fig:beam}). Fortunately silicon is tough enough that the risk of breakage is not an issue. The main limitation on the length of the device comes from buckling, just as for a straight shaft. Calculations show that an array with curvature radius of 5 mm can deliver insertion forces 4-fold higher than necessary (Section \ref{sec:failure}).

A different build option makes use of the recently developed polymer arrays \cite{luanUltraflexibleNanoelectronicProbes2017,chungHighDensityLongLastingMultiregion2019}. These devices are flexible, and consist of a thin and narrow strip of polymer that carries a string of recording sites and the associated wires. Obviously such a thread cannot be pushed into the brain. Instead it gets pulled into the tissue by a stiff needle that acts as a shuttle \cite{zhaoParallelMinimallyinvasiveImplantation2019,muskIntegratedBrainMachineInterface2019,thielenComparisonInsertionMethods2021}. When the needle withdraws, it leaves the polymer thread behind. Again, only straight needles have been used so far, but there is no reason for this constraint. One can construct a circular shuttle and insert it into the tissue by the same rotary motion discussed above (Fig \ref{fig:construction}b). The flexible thread will follow the shuttle on its trajectory, and when the shuttle withdraws (reversing the same trajectory), a circular electrode array is left behind. 

One advantage of the shuttle-and-thread option over the silicon design is that it completely separates the manufacture of the array from the choice of its shape. One can imagine a toolbox of reusable wire shuttles with different radii that each serve to implant polymer threads of varying designs. Also this method naturally extends to multiple polymer threads: a multi-shank "comb" of circular shuttles can simultaneously implant a whole family of electrode threads \cite{zhaoParallelMinimallyinvasiveImplantation2019}, effectively creating a curved surface of recording sites within the tissue. An advantage of the silicon design, at least for now, is that it offers more and denser electrode sites, and the option of electrode switching with circuits located on the array \cite{junFullyIntegratedSilicon2017}.

\section{Helical arrays}
\begin{figure}
  \centering
    \includegraphics[width=1.0\linewidth]{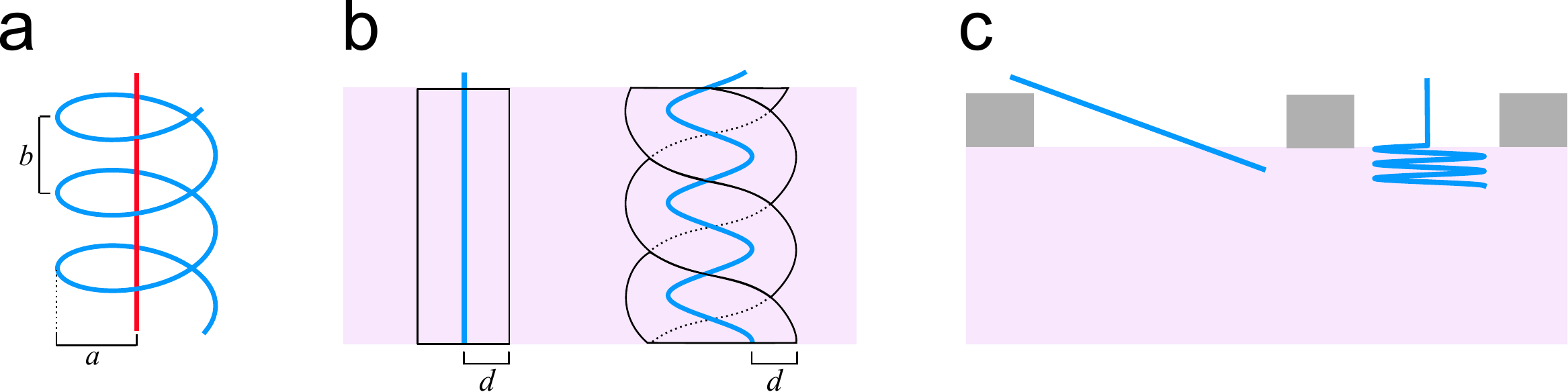}
    \caption{\textbf{Some uses of helical electrodes.} 
    \textbf{a:} A circular helix viewed at an angle of 30 degrees. Definitions of the radius $a$ (measured perpendicular to the red axis) and pitch $b$ (measured parallel to the axis). 
    \textbf{b:} A helical electrode covers greater volume. Suppose all neurons within distance $d$ of the array can be recorded. Left: A straight electrode array covers a cylinder of radius $d$. Right: An optimized helical array, with $a=d$ and $b=2d$ covers a compact volume 3.25 times larger. 
    \textbf{c:} Recording densely in the brain's superficial layer. Left: Conventional approach uses a straight array at a shallow approach angle, which requires a large hole in the skull (grey). Right: Helical array with a shallow angle $b/a$ can sample neurons densely even within a small hole.}
\label{fig:helical}
\end{figure}

As an electrode array (or shuttle) gets pushed into the brain, it must at all points travel tangentially to its own surface. Any sideways movement will slice apart the neural circuits that one would like to observe. The most general shape that satisfies this constraint is a circular helix, as illustrated by a corkscrew. Such a helix is defined by two parameters (Fig \ref{fig:helical}a): its radius (namely the radius of the tube that encloses it) and its pitch (the distance along the helix axis from one turn to the next). The circle ($\textrm{pitch}=0$) and the straight line ($\textrm{radius}=0$) are simply special cases of a helix.

To insert a helix into brain tissue, one needs to pair advancing motion with rotation. This can be implemented with a screw drive: If the pitch of the screw matches the pitch of the helix then advancement and rotation will be evenly matched. For manufacture of a helical array, silicon fabrication seems challenging, because the patterning no longer takes place in a plane. The shuttle-with-thread option doesn't suffer from this concern. A shuttle requires no fancy patterning and can be made, for example, from 12 µm tungsten wire \cite{zhaoParallelMinimallyinvasiveImplantation2019} bent into the desired helix. Some miniaturized coil springs made of tungsten wire are even available off-the-shelf and might serve as starting material to be thinned and sharpened by eletrolytic etching.

One use case for helical electrodes involves sampling all neurons within a compact column of the cortex. What would be the optimal helix for this purpose? Suppose that the electrodes can record every neuron within a distance $d$ of the polymer array. Then a single helical array with ${\rm{radius}} = d, {\rm{pitch}} = 2d$ should reach almost every neuron in a compact column (Fig \ref{fig:helical}b). That volume is 3.25 times larger than what can be accessed from a single straight array.  

Another application for helical electrodes arises if one wants to sample neurons densely at the surface of the brain, for example to inspect the arrangement of a sensory map, or assess the correlations between nearby cells in the same layer \cite{hubelFerrierLectureFunctional1977}. The conventional approach is to carve out a wider craniotomy so that a linear array can be inserted from a distance at a shallow angle (Fig \ref{fig:helical}c). Alternatively one might use a helical needle with a short pitch to implant a polymer array at that same shallow angle, but all within the confines of a small craniotomy. One could even install multiple threads using the same shuttle, taking care to use a different entry point into the brain each time so the threads don't interfere.

\section{How to explore the benefits of curved arrays}
Until the day comes when we can record from every neuron in the brain simultaneously \cite{kleinfeldCanOneConcurrently2019}, we still get to make choices about which neurons to monitor. The brain arranges its neurons in a great variety of structures: nuclei, glomeruli, barrels, hypercolumns, and layers of varying thickness and curvature. There is no reason to stick with a single tool shape to access all these diverse geometries. The present proposal expands the space of possible electrode shapes from a single option (straight line) to infinitely many (all possible helices). Particularly with a view towards implants in the human brain, the additional degrees of freedom should help in planning electrode trajectories that minimize collateral damage. 

I am hoping that readers will imagine new use cases and create a wish list for shaped electrode arrays that could fulfill their desires. For circular arrays in silicon, the technical hurdles for construction appear small. An open-source platform exists for manufacture of such arrays \cite{yangOpenSourceSilicon2020}: If several research groups join resources here, one could design a new mask with multiple prototype circular arrays at a modest cost to each user. With some collective initiative, exploring the benefits of curved electrode arrays seems well within reach, and I'll be happy to help coordinate such a foray if there is interest.

\newpage
\section{Appendix: Failure analysis for a curved silicon array} \label{sec:failure}
\begin{figure}
  \centering
    \includegraphics[width=0.5\linewidth]{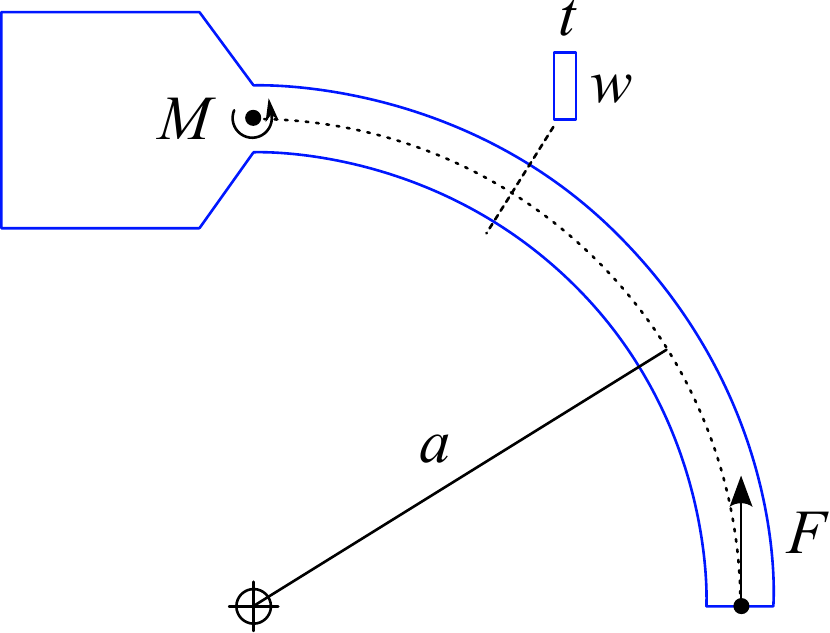}
    \caption{\textbf{Forces and moments on a curved beam of radius $a$, width $w$ and thickness $t$.} The force $F$ applied at the tip produces a maximal bending moment $M = F a$ at the base. If that moment is too large the beam breaks at the base.}   
\label{fig:beam}
\end{figure}

For a straight silicon array pushed into the brain, the main failure mode is buckling \cite{thielenComparisonInsertionMethods2021}. If the force needed to puncture and penetrate the brain surface exceeds the buckling force of the silicon shaft, the array will bend in half. This constraint ultimately limits the length of the shaft. For a curved silicon device, another failure mode appears, namely fracture at the base of the shaft (Fig \ref{fig:beam}). This happens when the bending moment $M = F a$ produces stress exceeding the failure limit. Here I explore the limits on the size of a curved array from both constraints.

\subsection{Fracture}

A circular silicon array can be treated as a curved beam with rectangular cross section (Fig \ref{fig:beam}). For concreteness I will use the dimensions of the Neuropixels 1.0 array \cite{junFullyIntegratedSilicon2017}, with a width in the plane of curvature of $w=70{\textrm{ µm}}$ and  thickness $t=20{\textrm{ µm}}$. The maximal stress at the base is

$$ S = 6 \frac{F a}{t w^2} $$

The fracture stress for silicon microdevices has been measured at 

$$ S_{\rm{frac}} \approx 1.5 \times 10^9 {\textrm{ Pa}} $$

Note this is about 5 times that of bulk silicon \cite{najafiStrengthCharacterizationSilicon1990}. Keeping the maximal stress below the failure limit one obtains a limiting force of 

$$ F_{\rm{frac}} = \frac{S_{\rm{frac}} t w^2}{6 a} = \frac{2.45 \times 10^{-5} {\textrm{ Nm}}}{a} $$

How much force is needed to push the electrode array into the brain? For Neuropixels the penetration force into mouse brain has been measured \cite{obaidUltrasensitiveMeasurementBrain2020} at 

$$ F_{\rm{pen}} = 0.3 {\textrm{ mN}} $$

Immediately after penetration of the pia that force drops to a much lower value. Putting it all together one gets an upper limit on the radius $a$ of the curved array:

$$ a_{\rm{frac}} = \frac{S_{\rm{frac}} t w^2}{6 F_{\rm{pen}}} = 82 {\textrm{ mm}} $$

This is much larger than any plausible application might call for. 

\subsection{Buckling}

The curved beam of Figure \ref{fig:beam} can also buckle by bending out of the plane of the beam. The boundary conditions are: At the left end the beam is fixed, and at the right end it is pinned but free to turn about all 3 axes. The analytical treatment of this elasticity problem looks hairy and is best left to professionals. Using finite element modeling I found that the buckling force for the curved beam of Fig \ref{fig:beam} is 0.237 times that for a straight beam of length $a$ (fixed at one end, pinned at the other), namely \cite{timoshenkoTheoryElasticStability1961}

$$F_{\rm{buck}} = 0.237 \frac{E t w^3}{a^2} \frac{\pi^2}{12 K^2} , K \approx 0.699$$

where

$$E = 1.6 \times 10^{11} {\textrm{ Pa}} $$

is the elastic modulus of silicon. Setting the buckling force to the penetration force one derives a maximal radius for the curved array of 

$$ a_{\rm{buck}} = \sqrt{\frac{E t w^3}{F_{\rm{pen}}}\frac{0.237 \pi^2}{12 K^2}} \approx 10.9 {\textrm{ mm}} $$

Clearly buckling is a more severe constraint than fracture, and it limits the plausible radius of curvature to 10 mm. 

\section{Acknowledgements}
Thanks to Kyu Hyun Lee, Sotiris Masmanidis, and Harri Kytomaa for feedback and advice. MM is supported in part by grants from the Simons Foundation (543015) and from the NINDS (5R01NS111477).

\newpage
\section*{References}
\renewcommand{\bibsection}{}\vspace{0em}
\bibliographystyle{unsrtnat}
\bibliography{references}

\end{document}